\documentclass[aps,prd,amsmath,nofootinbib,preprint,tightenlines,showpacs,superscriptaddress
]{revtex4}
\newcommand{\be}{\begin{equation}}
\newcommand{\ee}{\end{equation}}
\newcommand{\bea}{\begin{eqnarray}}
\newcommand{\eea}{\end{eqnarray}}
\newcommand{\bml}{\begin{subequations}}
\newcommand{\eml}{\end{subequations}}
\newcommand{\elea}{\eea\eml}
\newcommand{\blea}{\bml\bea}

\newcommand{\Vout}{V_{\text{out}}}
\newcommand{\Vin}{V_{\text{in}}}
\newcommand{\Kout}{K_{\text{out}}}
\newcommand{\Kin}{K_{\text{in}}}
\newcommand{\oin}{\omega_{\text{in}}}
\newcommand{\oobs}{\omega_{\text{obs}}}

\usepackage{epsfig}

\begin{document}

\title{Detectability of gravitational effects of supernova neutrino
  emission through pulsar timing}

\author{Ken D. Olum}
\author{Evan Pierce}
\affiliation{Institute of Cosmology, Department of Physics and Astronomy,\\ 
Tufts University, Medford, MA 02155, USA}
\author{Xavier Siemens}
\affiliation{Center for Gravitation and Cosmology, Department of
Physics, University of Wisconsin--Milwaukee,
P.O. Box 413, Milwaukee, Wisconsin, 53201}

\begin{abstract}

Core-collapse supernovae emit on the order of $3\times 10^{53}$ ergs
in high-energy neutrinos over a time of order 10 seconds, and so
decrease their mass by about 0.2 $M_\odot$.  If the explosion is
nearly spherically symmetric, there will be little gravitational wave
emission.  Nevertheless, the sudden decrease of mass of the progenitor
may cause a change in the gravitational time delay of signals from a
nearby pulsar.  We calculate the change in arrival times as successive
pulses pass through the neutrino shell at different times, and find
that the effect may be detectable in ideal circumstances.

\end{abstract}


\maketitle

\section{Introduction}

Core-collapse supernovae emit most of their energy in the form of
neutrinos, with on the order of $3\times 10^{53}$ ergs emitted.  Such
neutrinos were detected from SN1987A over the course of 13 seconds.
If the emission is anisotropic, these neutrinos could be the source of
a gravitational wave signal, but that will not be our interest here,
nor will we be concerned with any non-gravitational interactions of
the emitted neutrinos.  Instead we consider gravitational time-delay
effects resulting from the sudden expulsion of mass of order 0.2
$M_\odot$ into a shell that moves outward at the speed of light.

We propose to detect these effects via pulsar timing in the case where
there is a stable pulsar located near the supernova.  The magnitude of
the effect can be estimated as follows.  Suppose a pulsar and a
core-collapse supernova are are equidistant from Earth and separated
from each other by distance $b$.  Before the supernova, signals from
the pulsar are gravitationally redshifted by an amount $z_i =
GM_i/(c^3 b)$, where $M_i$ is the mass of the star before the
supernova, $G$ Newton's constant, and $c$ the speed of light.  We have
ignored effects of higher order in $GM/(b c^3)$.  Signals emitted from
the pulsar once time $b$ has elapsed after the supernova are
redshifted by the smaller amount $z_f = GM_f/(c^3 b)$, where $M_f$ is
the mass of the star excluding energy emitted at the speed of light,
which is primarily in the form of neutrinos.  Later signals thus are
blueshifted relative to those emitted before the supernova by
\be
z = Gm/(c^3 b),
\ee
where $m$ is the mass emitted in neutrinos.

If we have an observation period which is also equal to $b$, the
blueshift $z$ will cause an advance of the time of arrival by $Gm/c^3$
over the period of observation, with no dependence on $b$.  With $m
\sim 0.2 M_\odot$, $Gm/c^3 \sim 1\mu s$, and since pulsar timing accuracy
is at the level of tens of nanoseconds, such an effect could easily be
detected.

In the rest of this paper, we analyze this effect more carefully.
In Sec.~\ref{sec:metric} we give the effect of passing through the
shell on the frequency of a photon and the motion of an observer.  In
Sec.~\ref{sec:blueshift} we compute the resulting blueshift of the
pulsar signals as observed on Earth, in Sec.~\ref{sec:toa} we give the
resulting advancement of pulse arrival times, and in
Sec.~\ref{sec:observation} we discuss the prospects for observation.

\section{Neutrino shell metric}\label{sec:metric}

We will now compute the metric of the spacetime containing the
expanding neutrino shell.  We will not include the mass that stays
behind in the remnant, because it was present before and after the
supernova and thus does not contribute at first order to the change in
time delay.  We also ignore the ordinary matter ejected during the
explosion, because this moves at much less than the speed of light and
so will not reach distances comparable to $b$ during a reasonable
period of observation.  So we will take the spacetime inside the
expanding shell to be flat and spacetime outside to be given by the
Schwarzschild metric with mass $m$.

One could join these two regions by using the Israel junction
conditions \cite{Israel:1966rt}, but it seems to be somewhat easier
just to use the Vaidya metric \cite{Lindquist:1965zz} directly.  So let
$u$ be a null coordinate constant on outgoing rays and let $m(u)$ be
the mass inside coordinate $u$.  Let $u=0$ be the coordinate of the
neutrino shell, so that $m(u)$ decreases from $m$ to 0 in a narrow
range of $u$ around 0.

We then have the metric in spherical coordinates
\cite{Lindquist:1965zz},
\be
ds^2 = - f du^2 - 2 du dr + r^2 d \Omega^2,
\ee
where $d\Omega^2 =  d\theta^2 + \sin^2 \theta d\phi^2$,
\be
f \equiv 1 - \frac{2 m(u)}{r},
\ee
and we work in units where $c = G =  1$.

We can convert to a Schwarzschild-like time coordinate $t$ by taking
\be
u = t - r - 2 m \ln(r - 2m),
\ee
so
\be
 du = dt - \frac{dr}{f}.
\ee
Inside the shell we have flat space,
\be
ds^2 = - dt^2 + dr^2 + r^2 d \Omega^2,
\ee
and outside we have the Schwarzschild metric,
\be
ds^2 = -f dt^2+ \frac{dr^2}{f} + r^2 d \Omega^2.
\ee
Let $\Vout^a = (\Vout^t, \Vout^r, \Vout^\theta, \Vout^\phi)$ be some
vector outside the shell.  If we parallel transport $V^a$ along any
path $x(\lambda)$ that crosses the shell, we find
\be
\frac{d V^a}{d \lambda} = - \Gamma^a_{b c} \left( \frac{d x^b}{d \lambda} \right) V^c.
\ee
We can take a very short path across the shell, so that the connection
will only be important if it has components depending on $dm/du.$
There is just one such component, $\Gamma^r_{uu} = - \left( dm/du
\right)/r$.  So the only term in the parallel transport equation is
\be
\frac{d V^r}{d \lambda} = \frac{1}{r}\left( \frac{d m}{d u} \right)
\left( \frac{du}{d \lambda} \right) V^u = \left( \frac{d m}{d \lambda}
\right) \frac{V^u}{r}.
\ee
Thus the integral does not depend on the path, and crossing the shell
from outside ($m(u) = m$) to inside ($m(u) = 0$) decreases $V^r$ by
$(m/r) V^u$, so
\be
\Vin^r = \Vout^r - \frac{m}{r} \Vout^u,
\ee
and otherwise $\Vin^a = \Vout^a$.  We can convert to $(t,r)$
coordinates using
\be
V^u = V^t - \frac{V^r}{f}.
\ee
to find
\blea
\Vin^r &=&  \left( 1 + \frac{m}{r f} \right)\Vout^r - \frac{m}{r} \Vout^t,\\
\Vin^t &=&  \left( 1 - \frac{m}{r} \right)\Vout^t
 -  \frac{m}{r f}  \Vout^r.
\elea
We will ignore terms of order $m^2$ throughout, so we can write
\blea
\Vin^r &=&  \left( 1 + \frac{m}{r} \right)\Vout^r  - \frac{m}{r}\Vout^t,\\
\Vin^t &=&  \left( 1 - \frac{m}{r} \right) \Vout^t - \frac{m}{r}  \Vout^r.
\elea

\section{Blueshift From Shell}\label{sec:blueshift}

Now consider a photon that is emitted outside the shell but observed
on Earth after the shell has passed.  The mass of the shell leads to a
gradual deflection of the photon path while the photon is outside the
shell, and crossing through the shell leads to a sudden deflection.
But the size of these deflections is linear in $m$ and so their
effects on frequency and arrival time will be proportional to $m^2$,
and can be ignored.

The important effect is the difference in frequency of photons that
pass through the shell at different places.  Let $\omega_0$ be the
photon frequency measured by a stationary observer at infinity.  If
the photon crosses the shell at radius $r$, it will have immediately
before the crossing the frequency 4-vector
\be
\Kout^\mu = \left(f^{-1}\omega_0, K^r, K^\theta, K^\phi\right),
\ee
with $(K^r)^2 + f r^2 [(K^\theta)^2 + \sin^2\theta (K^\phi)^2] =
\omega_0$.

Inside the shell, the photon frequency becomes
\be
\Kin^0 = \omega_0 f^{-1} \left( 1 - \frac{m}{r}\right)
-  K^r\frac{m}{r}
= \omega_0 \left( 1 + \frac{m}{r}\right)-  K^r\frac{m}{r}.
\ee
If the photon makes angle $\alpha$ to the radial direction at the time
of crossing, we thus have
\be
\oin = \Kin^0 = \omega_0\left[1 + \frac{m}{r}(1- \cos\alpha)\right].
\ee
This corresponds to the photon having
received an inward boost with velocity $m/r$.  If the photon is
traveling radially inward, then $\oin = \omega_0 f_R^{-1}$.
For photons traveling nearly radially outward, and thus barely
overtaken by the shell, $\oin = \omega_0$.

One might think that the sudden frequency change on crossing the
shell would lead to a discontinuous frequency shift of observed
signals, and that such an effect could be observed even at great
distances from the supernova.  But this is not the case.  Consider an
observer who is stationary at fixed radius $r$ outside the shell.  His
4-velocity in $(t,r,\theta,\phi)$ coordinates is
\be
U_{\text{out}}^\mu = \left(f^{-1/2},0,0,0 \right).
\ee
If he measures the photon immediately before crossing the shell, he
finds frequency $\omega_0 f^{-1/2}$, which represents the blueshift
due to the mass of the shell.  If the observer moves inertially as the
shell passes, his 4-velocity afterwards is
\be
U_{\text{in}}^\mu = \left( 1, -\frac{m}{r},0,0 \right).
\ee
The photon crosses at the same place, and the observer's measurement
of the photon frequency is
\be\label{eqn:oinseen}
\oin + \frac{m}{r} K^r = \omega_0 f^{-1/2},
\ee
just as before.  The shell gives equal boosts to the observer and the
photon, so the observed frequency does not change.  The only
measurable effect is thus a smooth change in observed frequencies
depending on where and when different photons crossed through the
shell.

We expect the distance between Earth and the supernova to be so large
that the gravitational effect of the neutrino shell passing us is
negligible, so we will take the observed frequency to be just $\oin$.
When we observe a photon at time $t$, we find a frequency
\be
\oobs(t) = \omega_0\left[1 + \frac{m}{r(t)}(1- \cos\alpha(t))\right],
\ee
where $r(t)$ is the radial position at which the shell and the photon
cross, and $\alpha(t)$ is the angle between the radial direction and
the photon direction at that time.  We compute these quantities in the
next section.

\section{Time-shift of pulsar data from shell}\label{sec:toa}

\begin{figure}
\centering
\epsfxsize=3in\epsfbox{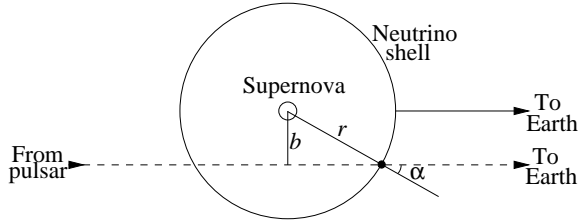}
\caption{Signals from a pulsar passing through a shell of neutrinos
  emitted by a supernova.  At the time of crossing, the radius of the
  shell is $r$ and the direction of signal propagation makes angle
  $\alpha$ to the radial direction.}
\label{fig:geometry}
\end{figure}

We will first consider a pulsar located very far behind the supernova.
We let $b$ be the distance of closest approach between the supernova
center and the photon trajectory, as shown in Fig.~\ref{fig:geometry},
and let $t=0$ denote the time when the neutrino shell reaches Earth.
We take the distance to Earth to be much larger than $b$ and we
continue to work at first order in $m$.  If a photon crosses the shell
at radial distance $r$, it will reach Earth at time
\be\label{eqn:tr}
t = r(1-\cos\alpha).
\ee
Using $b=r\sin\alpha$, we can solve for $r$ in terms of $t$,
\be
r(t) = \frac{t^2 + b^2}{2 t} .
\ee
From Eqs.~(\ref{eqn:oinseen})--(\ref{eqn:tr}), the observed frequency of
a photon arriving at time $t$ is thus \
\be
\oobs(t) = \omega_0\left(1+\frac{mt}{r^2}\right)
= \omega_0\left(1 + \frac{4 m t^3}{\left( t^2 + b^2 \right)^2} \right).
\ee

Correspondingly if the period between pulses observed before the
supernova is $T_0$, the period between pulses observed at time $t$ is
\be
T(t) = T_0\left(1 - \frac{4 m t^3}{\left( t^2 + b^2 \right)^2} \right).
\ee
The period begins to decrease at $t=0$, but only at third order.
It reaches a minimum at $t =\sqrt{3}b$, and then returns
asymptotically to $T_0$.

The time of arrival advancement at time $t$ is
\be\label{eqn:TOA}
\Delta t = \sum (T_0 - T) = \int_0^t \frac{(T_0 - T(t))d t}{T_0}
= 2 m \left[\ln(1+\tau^2)- \frac{\tau^2}{1 + \tau^2}\right],
\ee
where $\tau \equiv t/b$.  Reinserting units, we can write the
prefactor
\be
2m\frac{G}{c^3} \approx 10 \mu s \left( \frac{m }{M_\odot}\right).
\ee
Thus the total magnitude of this effect is easily large enough to be
detected.  Unfortunately, the timescale is quite long, being given by
the impact parameter $b$.  Thus observations are only possible when
the supernova and the pulsar (or the line of sight to the pulsar) are
quite nearby.  We discuss this in the next section.

\begin{figure}
\centering
\epsfxsize=3in\epsfbox{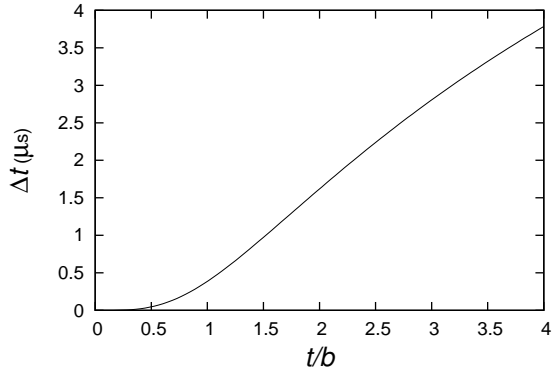}
\caption{Time-of-arrival advancement for a supernova emitting
  $0.2M_\odot$ of energy in a neutrino shell.}
\label{fig:dt}
\end{figure}

Equation~(\ref{eqn:TOA}) is plotted in Fig.~\ref{fig:dt}.
The time advancement increases very slowly at the beginning, so little
effect can be seen until about $t=b$.  There is an inflection point at
$t=\sqrt{3}b$, and eventually the increase is only logarithmic, but in
fact the curve is quite flat from $t\sim b$ to $t \sim 3b$.

Now consider the case where the pulsar is located near the supernova,
rather than far behind.  In this case we will see the time-of-arrival
curve exactly as for a distant pulsar, until the shell has passed the
pulsar.  After that, the pulse interval, and thus the slope of the
$\Delta t$ vs.\ $t$ curve, will remain the same.  This helps
observation if it happens when the slope is large, and hinders it if
the slope is small.  If the pulse emitted when the shell passes the
pulsar is received at time $t$, then the pulsar is further from Earth
than the supernova by distance $x=(t^2-b^2)/(2t)$.  The best possible
geometry has this happen at $t=\sqrt{3}b$, when the slope is greatest,
which gives $x=b/\sqrt{3}$.  But we can see from Fig.~\ref{fig:dt}
that the effect of the finite distance to the pulsar is unimportant as
long as the pulsar is not too far in front of the supernova.

\section{Prospects for observation}\label{sec:observation}

\begin{figure}
\centering
\epsfxsize=4in\epsfbox{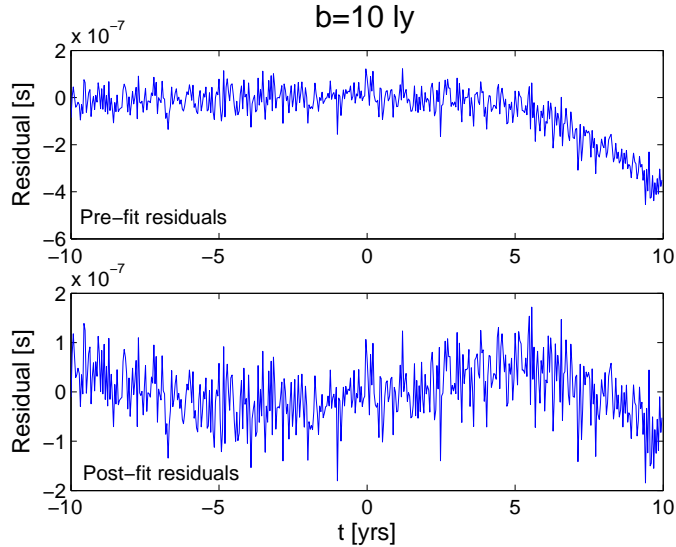}
\caption{Time-of-arrival residuals before (top panel) and after
  (bottom panel) subtraction of a quadratic fit.  We take the pulsar
  to have been observed for 10 years before and 10 years after the
  supernova was seen.  We have added 50ns of white noise to account
  for observational uncertainties.  Subtraction of the quadratic fit
  reduces the magnitude of the effect of the supernova relative to the
  noise, but it is still clearly visible.}
\label{fig:b10}
\end{figure}

Timing of pulsars requires modeling the unknown distance to the
source, pulse interval, and spin-down rate.  (One must also subtract
the known motion of the observing station relative to the sun, and the
periodic motion of the pulsar if it is in a binary.)  Thus one must
make a quadratic fit to observed data, and only deviations from this
fit can be detected.  In Fig.~\ref{fig:b10} we show potential data
from a scenario with $b = 10$ light-years.
In this case the effect is clearly observable, even after the best
quadratic fit has been subtracted.  The effect comes primarily from
the time-of-arrival advancement when $t\sim b$.  Thus if the the
period of observation after the supernova were significantly smaller
than $b$, the effect would not be observable.

A chance alignment of of a background pulsar with a nearby supernova
is quite unlikely.  Even if we were to be lucky enough to see a
supernova in our own galaxy, the typical distance would be of order
the distance to the galactic center, about 8kpc.  A disk of radius 10
light-years (3pc) at this distance would subtend solid angle only
$4\times 10^{-7}$ steradians, and thus the chance that the line of
sight to any given pulsar would pass through this disk is only
$3\times 10^{-8}$.

A better possibility is that the supernova and the pulsar are both
part of the same compact star-forming region.  For example, consider
the R136 region in the Large Magellanic Cloud.  This region produced a
supernova seen by John Herschel in 1836, and also contains a pulsar,
PSR J0537-6910, which was formed in a supernova about 4000 years ago.
The angular distance between these objects is $21''$.  Taking a
distance of 50kpc to R136 gives a transverse distance of only 16
light-years between the supernova and the pulsar.

SN 1987A occurred at the edge of the Tarantula Nebula, about 700 light
years from PSR J0537-6910.  The time-of-arrival advancement due to
this supernova is thus still in the initial quartic segment and would
not be detectable within a reasonable period of observation.

A supernova in a binary star system that already contains a pulsar
would be an even better but far rarer target.  Signals from the pulsar
that pass through the neutrino shell reach Earth before signals
emitted after the shell has reached the pulsar.  Thus the effect could
in principle be observed before any disturbance of the pulsar due to
its proximity to the supernova.

\section*{Acknowledgments}

We would like to thank Ben Shlaer and Carrie Thomas for useful
conversations.  This work was supported in part by the National
Science Foundation under grant numbers 0855447 and 1213888.

\bibliography{paper}

\begin{thebibliography}{2}
\expandafter\ifx\csname natexlab\endcsname\relax\def\natexlab#1{#1}\fi
\expandafter\ifx\csname bibnamefont\endcsname\relax
  \def\bibnamefont#1{#1}\fi
\expandafter\ifx\csname bibfnamefont\endcsname\relax
  \def\bibfnamefont#1{#1}\fi
\expandafter\ifx\csname citenamefont\endcsname\relax
  \def\citenamefont#1{#1}\fi
\expandafter\ifx\csname url\endcsname\relax
  \def\url#1{\texttt{#1}}\fi
\expandafter\ifx\csname urlprefix\endcsname\relax\def\urlprefix{URL }\fi
\providecommand{\bibinfo}[2]{#2}
\providecommand{\eprint}[2][]{\url{#2}}

\bibitem[{\citenamefont{Israel}(1966)}]{Israel:1966rt}
\bibinfo{author}{\bibfnamefont{W.}~\bibnamefont{Israel}},
  \bibinfo{journal}{Nuovo Cim.} \textbf{\bibinfo{volume}{B44S10}},
  \bibinfo{pages}{1} (\bibinfo{year}{1966}).

\bibitem[{\citenamefont{Lindquist et~al.}(1965)\citenamefont{Lindquist,
  Schwartz, and Misner}}]{Lindquist:1965zz}
\bibinfo{author}{\bibfnamefont{R.}~\bibnamefont{Lindquist}},
  \bibinfo{author}{\bibfnamefont{R.}~\bibnamefont{Schwartz}}, \bibnamefont{and}
  \bibinfo{author}{\bibfnamefont{C.}~\bibnamefont{Misner}},
  \bibinfo{journal}{Phys.Rev.} \textbf{\bibinfo{volume}{137}},
  \bibinfo{pages}{B1364} (\bibinfo{year}{1965}).

\end{thebibliography}

\end{document}